\documentstyle[12pt]{article}
\input{epsf}
\begin{document}
\addtolength{\baselineskip}{.25mm}
\newlength{\extraspace}
\setlength{\extraspace}{2.5mm}
\newlength{\extraspaces}
\setlength{\extraspaces}{2.5mm}

\newcommand{\newsection}[1]{
\vspace{15mm}
\pagebreak[3]
\addtocounter{section}{1}
\setcounter{subsection}{0}
\setcounter{footnote}{0}
\noindent
{\Large\bf \thesection. #1}
\nopagebreak
\medskip
\nopagebreak}

\newcommand{\newsubsection}[1]{
\vspace{1cm}
\pagebreak[3]
\addtocounter{subsection}{1}
\addcontentsline{toc}{subsection}{\protect
\numberline{\arabic{section}.\arabic{subsection}}{#1}}
\noindent{\large\bf 
\thesubsection.
#1}
\nopagebreak
\vspace{2mm}
\nopagebreak}
\newcommand{\ba}{\begin{eqnarray}
\addtolength{\abovedisplayskip}{\extraspaces}
\addtolength{\belowdisplayskip}{\extraspaces}
\addtolength{\abovedisplayshortskip}{\extraspace}
\addtolength{\belowdisplayshortskip}{\extraspace}}
\newcommand{\zbar}{\overline{z}}
\newcommand{\tr}{{\rm tr\,}}
\newcommand{\ea}{\end{eqnarray}}
\newcommand{\OL}[1]{ \hspace{2pt}\overline{\hspace{-2pt}#1
   \hspace{-1pt}}\hspace{1pt} }
\newcommand{\is}{& \! = \! &}
\newcommand{\eps}{\epsilon}
\newcommand{\calR}{{\cal R}}
\newcommand{\calB}{{\cal M}}
\newcommand{\calK}{{\cal K}}
\newcommand{\calG}{{\cal G}}
\newcommand{\tilF}{{\tilde F}}
\newcommand{\barG}{{\OL G}}
\newcommand{\alphap}{}
\newcommand{\be}{\begin{equation}
\addtolength{\abovedisplayskip}{\extraspaces}
\addtolength{\belowdisplayskip}{\extraspaces}
\addtolength{\abovedisplayshortskip}{\extraspace}
\addtolength{\belowdisplayshortskip}{\extraspace}}
\newcommand{\ee}{\end{equation}}
\newcommand{\STr}{{\rm STr}}
\newcommand{\figuur}[3]{
\begin{figure}[t]\begin{center}
\leavevmode\hbox{\epsfxsize=#2 \epsffile{#1.eps}}\\[3mm]
\parbox{15.5cm}{\small 
\it #3}
\end{center}
\end{figure}}
\newcommand{\im}{{\rm Im}}
\newcommand{\calm}{{\cal M}}
\newcommand{\call}{{\cal L}}
\newcommand{\sect}[1]{\section{#1}}
\newcommand\hi{{\rm i}}

\begin{titlepage}
\begin{center}

{\hbox to\hsize{
\hfill SU-ITP-01/50,~SLAC-PUB-9083}}
{\hbox to\hsize{
\hfill PUPT-2019}}

{\hbox to\hsize{
\hfill hep-th/0112197}}

\vspace{3.5cm}

{\large \bf Brane/Flux Annihilation and the String Dual\\[6mm]
of a Non-Supersymmetric Field Theory}\\[1.5cm]

{Shamit Kachru${}^2$,\ John Pearson${}^1$ and Herman Verlinde${}^1$}\\[8mm]

{${}^1$ \it Department of Physics, Princeton University, Princeton,
NJ 08544}\\[6mm]

{${}^2$ \it Department of Physics and SLAC, Stanford University, Stanford, CA
94305/94309}

\vspace*{3.5cm}

{\bf Abstract}\\

\end{center}
\noindent
We consider the dynamics of $p$ anti-D3 branes inside the Klebanov-Strassler
geometry, the deformed conifold with $M$ units of RR 3-form flux around the
$S^3$. We find that for $p<\!<M$ the system relaxes to a nonsupersymmetric
NS 5-brane ``giant graviton'' configuration, which is classically stable, but
quantum mechanically can tunnel to a nearby supersymmetric vacuum with
$M-p$\ D3 branes. This decay mode is exponentially suppressed and
proceeds via the nucleation of an NS 5-brane bubble wall. We propose a dual
field theory interpretation of the decay as the transition between a
nonsupersymmetric ``baryonic'' branch and  a supersymmetric
``mesonic'' branch
of the corresponding $SU(2M\!-p)\times SU(M\!-p)$ low energy gauge theory.
The NS 5-brane tunneling process also provides a simple visualization of
the geometric transition by which D3-branes can dissolve into 3-form
flux.

\end{titlepage}

\newpage

\sect{Introduction}

The search for calculable and nontrivial nonsupersymmetric string
vacua remains a problem of basic interest in string theory.
In the context of
AdS/CFT duality \cite{AdS}, this is related to the construction of
gravitational ``anti-holographic'' descriptions for nonsupersymmetric
4d quantum gauge theories.   Here, we describe the construction of
a simple nonsupersymmetric string background which has both
a holographic description in terms of 4d gauge
theory and realization in the context of warped
string compactifications.  The model possesses several attractive features,
among them the existence of certain discrete parameters which allow it to
be rendered classically stable and arbitrarily long-lived and the ability
to set the regime of nonsupersymmetric dynamics at energies exponentially
far below the Planck scale.

One of the recent advances in the anti-holographic description of
gauge theories is the discovery of several smooth gravitational duals for
${\cal N}\!=\!1$ supersymmetric pure Yang-Mills theory \cite{KS,MaNu,Vafa}.
One such
dual, as discussed by Klebanov and Strassler (KS), starts with the
theory of $N$ D3 branes probing $M$ ``fractional'' wrapped D5 branes in
a conifold geometry.  In the final geometry, the branes have disappeared;
they are replaced by
RR and NS three-form fluxes through two
intersecting three-cycles in the deformed conifold \cite{KS}.
The crossed fluxes carry $N$ units of $F_{\it 5}$-form charge, in keeping
with the starting point. The holographic description of
the geometry involves a 4d $SU(N+M) \times SU(N)$
gauge theory undergoing an RG cascade,
in which repeated Seiberg duality transformations relate weakly-coupled
descriptions of the physics. In the case
$N = KM$ for some integral $K$, the IR field theory at the
end of the cascade is a pure ${\cal N}\! =\! 1$ supersymmetric
$SU(M)$ Yang-Mills theory. 
The KS model can be embedded into a
compact geometry \cite{fluxhier} and used to realize warping as a
means of generating large hierarchies as in the Randall-Sundrum
scenario \cite{RS} along the lines advocated in \cite{HolComp} (for
other discussions along these lines, see e.g. \cite{Mayr}).
Roughly speaking, the pure Yang-Mills theory
becomes the theory on the ``infrared brane'' in the  RS description,
while the string compactification manifold provides an effective
``Planck brane".

In our model, along with the conifold geometry with crossed three-form
fluxes, we introduce an additional ingredient which breaks the supersymmetry
of the solution: We place some number $p <\!< M$
of anti-D3 branes at the tip of the deformed conifold.
We expect that this configuration can be classically stable
since, although the fluxes carry D3 brane charge, there is no immediate
analogue of brane/anti-brane annihilation.  Instead, the system must
relax by undergoing some form of ``brane/flux annihilation'' in which
one of the 3-form fluxes jumps by one unit, allowing part of the D3-brane
charge to materialize in the form of actual branes. Here we will identify
and analyze this process in some detail; we will find that, in the regime
of interest to us, it is an exponentially suppressed tunneling effect
which proceeds via the nucleation of an NS 5-brane bubble wall.
We will argue that, in the holographic dual description,
the decay represents a transition between a metastable
non-supersymmetric ``baryonic'' branch and  a supersymmetric
``mesonic'' branch of the corresponding $SU(2M-p)\times SU(M-p)$ low
energy gauge theory.

The organization of this paper is as follows.  In \S2, we discuss
preliminaries and describe our basic physical expectations.  We also
discuss the two different vacuum branches of the supersymmetric KS system,
since this structure will be of importance to us later. Then, in
sections \S3 and \S4, we study the basic dynamics of the anti-D3 branes
in the conifold and flux background.
We will do this from two perspectives.  In \S3, we will look at the
flux induced potential for anti-D3 branes and present evidence that it
induces them to expand into an NS 5-brane giant graviton configuration.
In \S4\ we will then examine this system directly from the viewpoint of
the NS 5-brane action with induced anti-D3 charge. By combining both
perspectives and establishing a precise correspondence between them,
we obtain a consistent picture where the anti-branes blow up into a fivebrane
wrapping an $S^2$ of some fixed size on the small (A-cycle) $S^3$ inside
the conifold. For small enough $p/M$, this describes a metastable
false vacuum. In \S4.3\ we present the analysis of the Euclidean NS 5 brane
bubble wall which mediates the quantum decay of our theory to a nearby
supersymmetric vacuum with slightly different fluxes and $M-p$ D3 branes.
This is the true vacuum, and is one of the models discussed in \cite{KS}.
In \S5\ we present our best guess for the correct holographic
description of our non-supersymmetric theory in terms of a
long-lived false vacuum in the KS field theory. We also discuss the
type IIA dual description. Finally, in 
section \S6, we give a short discussion of the
model when considered as a warped string compactification and describe
some features of the 4d effective supergravity theory.

As an interesting aside, we note that our study also sheds light on the
geometric transition \cite{Vafa} between string backgrounds with
D3-branes sources and smooth KS-geometries with only 3-form flux
\cite{KS}. Contrary to expectation,
we will see that this transition can be understood in terms of the
dynamics of the D3-branes themselves, which undergo
via the formation of
a ``fuzzy'' NS 5-brane a continuous transmutation into
pure flux. This transition takes place inside
supersymmetric domain walls interpolating between the two phases,
the BPS domain wall solution described in \S4.2.


\sect{Preliminaries}

In this section, we review some of the relevant aspects of
the KS geometry and its holographically dual gauge theory description.
We describe our non-supersymmetric model and some of our
physical expectations in \S2.2.

\newsubsection{The Klebanov-Strassler geometry from F-theory}

In most of our investigation, we will consider the
non-compact limit of the Klebanov-Strassler geometry.
For purposes of presentation and to facilitate later discussions
of our model in the
context of warped string compactifications, we will start from
the point of view of F-theory with non-zero 3-form fluxes
$H_{\it 3}$ and $F_{\it 3}$.
Such models have been investigated from an M-theory and F-theory
perspective in, e.g.,\cite{Beckers,fcomp,fluxhier,curio}.
To simplify notation, we assume as in \cite{fluxhier} that we are
working in a limit where the F-theory compactification on
the fourfold $X$ can be viewed
as an orientifold of IIB string theory on a Calabi-Yau threefold $Y$.
Besides the 3-form flux, the geometry will typically
involve the insertion of $N_3$ D3-branes and/or
$\overline{N}_3$ anti-D3 branes. The net D3 charge
\be
Q_{3} = N_3 - \overline{N}_3
\ee
is fixed via the global tadpole condition
\begin{equation}
{\chi(X)\over 24} ~=~  Q_{3}
+ {1\over{2\kappa_{10}^2 T_3}}\int_{Y}
H_{\it 3} \wedge F_{\it 3}\ .
\label{tadpoleF}
\end{equation}
Here $\chi(X)$ is the Euler characteristic of the CY fourfold that
specifies the F-theory compactification, and $T_3$ is the D3-brane tension.

We are interested in expanding about a singular point in moduli space
where $Y$ has developed a conifold singularity.
The KS solution corresponds to the placing of $M$ units of $F_{\it 3}$
flux through the $A$-cycle of the conifold
and $K$ units of $H_{\it 3}$ through the dual $B$-cycle:
\ba
\frac{1}{4\pi^2 \alphap}
\int_{A} F_{\it 3} \is M\ ,\nonumber \\
\frac{1}{4\pi^2 \alphap} \int_{B} H_{\it 3} \is -K\ .
\label{ks}
\ea
Assuming no other (crossed) fluxes are present, we could in principle
choose $M$ and $K$ such that ${\chi\over 24} = MK$, so that D3 charge
conservation is satisfied without the need of extra D3-brane
insertions. We can therefore expect to find a smooth compactification geometry.
Indeed, as discussed in detail in \cite{fluxhier}, these fluxes give rise
to a non-trivial superpotential for the complex structure modulus
$z_A = \int_A \Omega$ that controls the relative size of the A-cycle,
stabilizing it at an exponentially small value
\be
\label{hier}
z_A \sim \exp(-2\pi K/Mg_{\rm s}).
\ee
As a result, the compactification geometry develops a smooth conical region,
described by the warped geometry of the deformed conifold
\begin{equation} \label{dconifold}
\sum_{i=1}^4 z_i^2 =
\varepsilon^2\ ,
\end{equation}
with $\varepsilon \sim z_A$ exponentially small.

The exponential hierarchy (\ref{hier}) has a natural explanation from
the holographic perspective as follows \cite{KS,fluxhier}:  The
geometry (\ref{dconifold}) is dual to an $SU(N+M) \times SU(N)$ gauge
theory, which has a non-trivial $\beta$-function proportional to $g_s
M$. As a result, its renormalization group flow towards the infrared
involves a cascade of successive strong coupling transitions
at successive scales $\mu_n$ with $\log(\mu_{n}/\mu_{n+1}) \sim
(g_s M)^{-1}$. Each strong coupling transition involves a Seiberg
duality transformation which lowers the rank of the larger of the two
gauge groups by $2M$. Assuming that $N=KM$, the maximal number of such
dualities is $K$, leaving in the IR a pure supersymmetric $SU(M)$
Yang-Mills theory. The total range of scales traversed in this
duality cascade is therefore $\log(\mu_0/\mu_K) \sim K/g_sM$,
consistent with the supergravity prediction (\ref{hier}).

\figuur{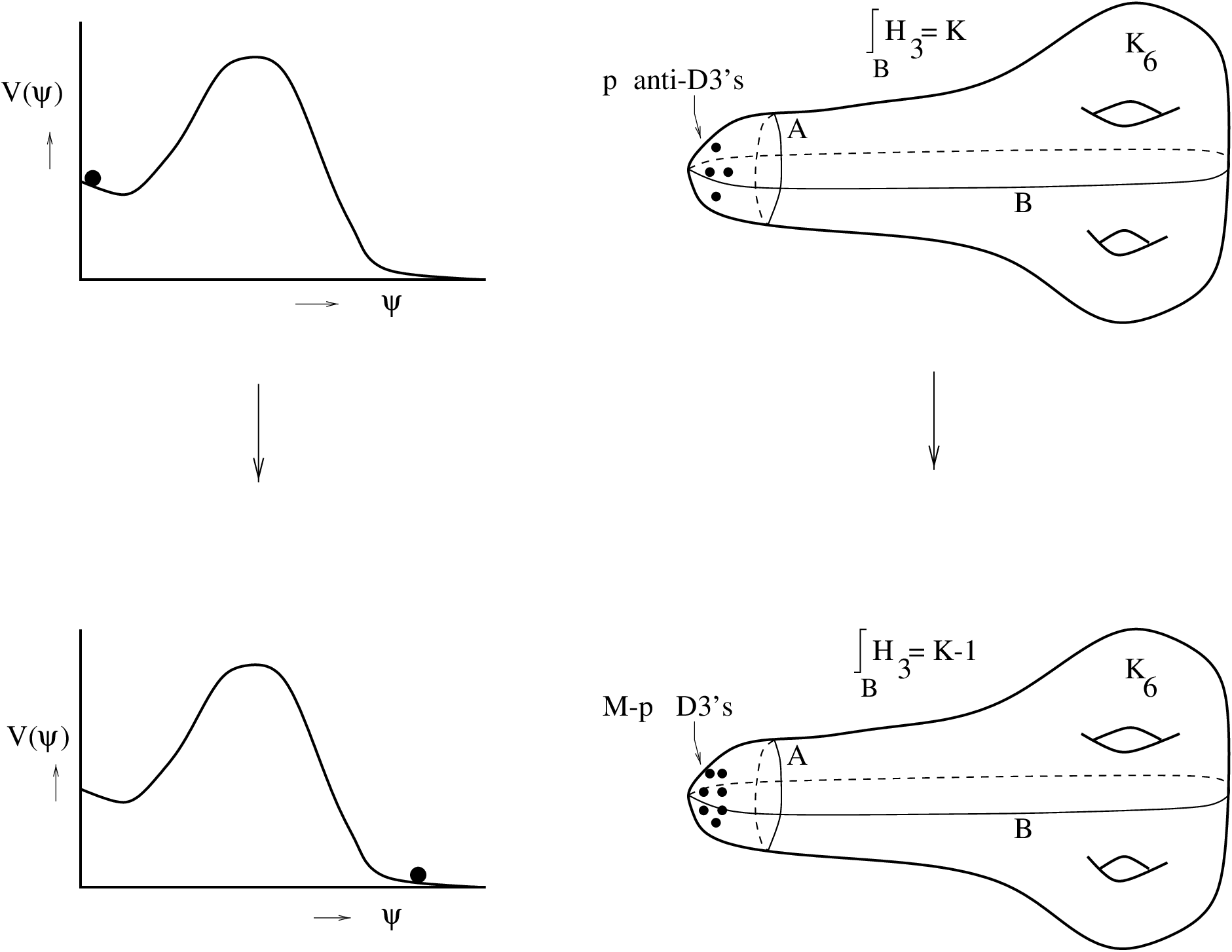}{10cm}{{\bf Fig 1:}The decay, described in \S2.2 and \S4.3, takes place between
an intitial non-supersymmetric situation with $p$ ${\overline{D3}}$ branes near the
tip of the conifold, to a final supersymmetric situation with $M-p$ D3-branes.
The total D3-charge is preserved via the simultaneous jump in the $H_3$
flux around the B-cycle by one unit, from $K$ to $K-1$.}

\newsubsection{Our Model and Physical Expectations}

We choose to study the model of \S2.1\ in the case that
\be
\label{ourcase}
N_3=0, ~{\overline N_3}=p,~{\chi\over 24} ~=~KM - p
\ee
with $p <\! < K,M$.  That is, we study the theory of $p$
anti-D3 branes probing the Klebanov-Strassler geometry.

This theory is nonsupersymmetric because the supersymmetry
preserved by the anti-D3 branes is incompatible with the global
supersymmetry preserved by the imaginary self-dual 3-form flux
of the background geometry. Thus, on general grounds, we should
expect this configuration to be unstable. Indeed, if the $KM$ units
of D3 charge contained in the $F_{\it 3}$ and $H_{\it 3}$ fluxes
were realized instead as real D3 branes, we would know that the system 
could decay back to a supersymmetric KS
vacuum by undergoing brane/anti-brane annihilation.  In our configuration,
however, there is no obvious way for this to occur.

To gain more insight, let us look at our situation (\ref{ourcase})
from the dual field theory perspective. It is natural to identify
the total D3 charge carried by the compactification manifold,
$\chi/24$, with the total rank $N$ of the $SU(N+M)\times SU(N)$
dual gauge theory. Indeed, if we took $N=KM + p$, with
$p < \! < M$, the duality cascade would proceed in $K$ steps until
we were left in the infrared with an $SU(M+p) \times SU(p)$ gauge theory. As
explained in \cite{KS}, this theory has a moduli space parametrized
by ``meson'' fields $N_{ij}$ which can be convincingly matched
with the moduli space of $p$ D3 branes moving on the deformed conifold
geometry. This is what we expect in the
case ${\chi\over 24} = KM+p$.

Extrapolating this correspondence, we might venture that our
situation (\ref{ourcase}) should be related to the KS gauge
theory with $N = KM-p$. However, in this case the RG cascade
can only involve $K-1$ Seiberg dualities, leaving a supersymmetric
$SU(2M-p)\times SU(M-p)$ gauge theory in the infrared.
Clearly, this field theory does not correspond to the non-supersymmetric
situation (\ref{ourcase}), but instead to the supergravity solution
with $M\!-\!p$ D3 branes and only $K\!-\!1$ units of $H_{NS}$ flux through
the B-cycle
\be
\label{finalstate}
N = M-p,~{\overline N}=0,~~{\chi\over 24} ~=~(K-1)M + (M-p) ~=~KM - p
\ee

It is a natural guess that this supersymmetric situation (\ref{finalstate})
should form the end point of a decay process beginning from the initial
situation (\ref{ourcase}). Indeed, one can see that the  IIB string theory
admits domain walls across
which the flux of $H_{\it 3}$ through the B-cycle drops by one unit,
described by NS 5 branes wrapping the A-cycle of the conifold geometry.
The system (\ref{ourcase}) can therefore decay via nucleation of a
bubble of supersymmetric vacuum (\ref{finalstate}) surrounded by a
spherical NS 5 domain wall. This domain wall will expand exponentially
as a consequence of the pressure produced by the drop in the vacuum
energy.  We will describe this process in much more detail
in \S4.3. We will find that it is a non-perturbative tunneling
process for a wide range of parameters.
However, this does not mean that the naive ${\overline{D3}}$ description of
the system is the correct description before the decay.
As we will see, the ${\overline{D3}}$
system also has ${\it perturbative}$ instabilities, which cause the
${\overline{D3}}$ branes to blow up into NS 5 branes wrapping spheres
inside the A-cycle. Provided $p$ is sufficiently smaller than $M$,
the maximum size ``giant graviton'' 5-brane will be a classically
stable configuration. Quantum mechanically it will decay with an
exponentially-suppressed probability.

To simplify our analysis, we will assume that all of the interesting
dynamics takes place very close to the tip of the conifold. 
Indeed, it is not difficult to convince oneself that the
anti-D3 branes will in general feel a net radial force $F_r(r)$
proportional to the 5-form flux $F_5$
\be
\label{force}
{F}_r(r) = - 2\mu_3 F_5(r)
\ee
that will attract them to the tip at $r=0$. This force
is a sum of gravitational and 5-form contributions:
For a $D3$-brane these two terms cancel; in the case of
$\overline{D3}$-branes they add up to a net attractive force
(\ref{force}). Therefore, even if we began with a more generic initial
distribution, we expect the anti-D3 branes to 
accumulate quickly near the apex at $r=0$.

The metric near the apex reads \cite{remarks}
(we work in string units, $\alpha'=1$)
\ba \label{apex}
ds^2  \is
 a_0^2\;
dx_\mu dx_\mu
+  \, g_s M\, b_0^2 
\Bigl( {1\over 2} dr ^2  + d\Omega_3^2 + r^2 d\tilde{\Omega}_2^2\Bigr)
\nonumber \\[2.5mm]
a_0^2 & \simeq &  { \varepsilon^{4/3}\over  g_s M\alphap }\,
\qquad \qquad b_0^2 \approx 0.93266.
\ea
Since we assume that all physics takes place at $r=0$,
our space-time has the topology ${\bf R}^4 \times S^3$.
The RR field $F_3 = dC_2$ has a quantized flux around the $S^3$
\be
\label{cflux}
\int_{S^3} F_3 = 4\pi^2 \alphap M,
\ee
while in the supersymmetric background dictates $\star_6 H_3 = -g_s F_3$,
so that
\be
\label{bsix}
dB_{\it 6} = {1\over g_s^2} \star_{10} H_{\it 3} =  -{1\over g_s} \,
dV_4 \wedge F_{\it 3},
\ee
with $dV_4 = \; a_0^4 \;  d^4 x$. The dilaton field is constant, and the
self-dual 5-form field vanishes at the tip.

\newsubsection{Branches of Moduli Space in the KS system}

Here we review one of the relevant features of the KS low energy field
theory, following section 7 of \cite{KS}.
This will be important in understanding our proposal for the holographic
field theory description of the metastable false vacuum, as well as of
the tunneling process that describes its quantum decay.

Consider the RG cascade of the KS gauge theory with $N=KM$ in its 
penultimate step.  In this case,
the unbroken gauge group is $SU(2M) \times SU(M)$, and the $SU(2M)$ factor
has an equal number of flavors and colors.
There is also a quartic superpotential whose rough form is
\be
\label{thesup}
W = \lambda \left(A_i B_j A_k B_l\right)\epsilon^{ik} \epsilon^{jl}
\ee
where the $A_i$ are in the $(2M,{\overline{M}})$ representation,
the $B_i$ are in the conjugate representation, and $i=1,2$.
We know from the analysis of \cite{Seiberg} that in ${\cal N}=1$
supersymmetric QCD with $N_f = N_c$, the moduli space is quantum mechanically
modified.  Treating the $SU(M)$ as a global symmetry (i.e. taking its
dynamical scale to vanish), this situation reduces to the one studied
in \cite{Seiberg} (with the added complication of the
quartic tree-level superpotential).

Define the ``meson'' fields $N_{ij,\alpha\beta}$ and the ``baryon''
fields ${\cal B}, \tilde{\cal B}$
\be
\label{mbdef}
N_{ij,\alpha\beta} ~=~A_{i,\alpha} B_{j,\beta},~~{\cal B}~=~(A_1)^M (A_2)^M,~~{\tilde{\cal B}} ~=~
(B_1)^M (B_2)^M\ ,
\ee
where $\alpha$ and $\beta$ are $SU(M)$ ``flavor'' indices.
In order to reproduce the quantum modified moduli space
of the $SU(2M)$ theory, we should add a Lagrange multiplier term to
(\ref{thesup})
\be
\label{modsup}
W = \lambda \left(N_{ij,\alpha\beta}
N_{kl}^{\alpha\beta} \right) \epsilon^{ik} \epsilon^{jl}
+ X (\det(N) ~-~{\cal B}{\tilde{\cal B}} - \Lambda^{4M})\ ,
\ee
where the determinant is understood to be that of a $2M \times 2M$ matrix (coming
from the $i,j$ and color indices on $N$).

There are distinct ``mesonic'' and ``baryonic'' branches of
supersymmetric vacua arising from the superpotential (\ref{modsup}).
On the baryonic branch
\be
\label{bary}
X~=~N~=~0,~~{\cal B}~=~{\tilde{\cal B}}~=~i\Lambda^{2M}.
\ee
On this branch, the $SU(M)$ factor in the gauge group remains unbroken,
and one is left with pure ${\cal N}=1$ gauge theory in the IR.
However, there is also a branch where ${\cal B} = {\tilde{\cal B}} = 0$
and the mesons have non-vanishing VEVs.
\be
\label{meso}
\det(N)~=~\Lambda^{4M},~~~{\cal B}~=~{\tilde{\cal B}}~=~0\ .
\ee
Now
the gauge group is generically Higgsed, and there is a moduli space of
vacua consisting of the theory of $M$ D3 branes probing a deformed
conifold geometry.

The fact that the mesonic and baryonic branches of moduli space are
disconnected is a result of the
tree level superpotential (\ref{thesup}).  In the theory without
(\ref{thesup}), the quantum moduli space is defined by the equation
$\det(N) -{\cal B} {\tilde{\cal B}} =\Lambda^{4M}$
and one can smoothly interpolate between these branches, via a continuous
path of supersymmetric vacua. In the present case, on the other hand, field
configurations that interpolate between the two branches take the form of
localized supersymmetric domain walls with non-zero energy density.
These domain walls solve BPS equations of the form
\be
\partial_z \Phi^I = g^{IJ}{\partial_J W(\Phi)}
\ee
with $\Phi^I =\{N_{ij}^{\alpha\beta},{\cal B}, X \}$, and $z$
the coordinate transverse to the domain wall. The wall tension is
proportional to
\be
|\Delta W|\, =\, |W(\Phi_b) - W(\Phi_m)|,
\ee
where $\Phi_b$ and $\Phi_m$ denote the respective vacuum values (\ref{bary})
and (\ref{meso}) of the two branches. By considering e.g. the special mesonic
vacuum where all meson fields have the same expectation value (i.e. independent
of their ``flavor'' index), we deduce
\be
|\Delta W|\, = \, 2\lambda M \Lambda^4.
\label{tension1}
\ee

In \S4.2 we will consider the dual supergravity description of the domain
wall. In this case it will represent the transition from the smooth
deformed conifold with only flux (corresponding to the baryonic branch)
to the situation with one less unit of NS 3-form flux and $M$ D3-branes.
In \S5 we will use this dual understanding of the supersymmetric
domain wall to motivate a similar holographic description of our
non-supersymmetric ${\overline{D3}}$ background and its decay process.
Here we will similarly argue that there are two relevant branches:
The first, the analogue of the baryonic branch, will be the (now
nonsupersymmetric) metastable vacuum; and the second, the analogue of
the mesonic branch, will be the theory with $M\! -\! p$
D3 branes probing the deformed conifold (\ref{finalstate}).

\newcommand{\Tr}{{\rm Tr}}
\newsection{The ${\overline{D3}}$ brane Perspective}

We are interested in the dynamics of $p$ ${\overline{D3}}$ branes
sitting at the
end of the KS throat, under the influence of the $F_{\it 3}$ and
$H_{\it 3}$ fluxes (\ref{ks}).
We will do our analysis within the probe approximation, taking the
KS background as fixed, while ignoring the backreaction due to the
${\overline{D3}}$ branes. The characteristic size of the geometry is set by
$R \simeq \sqrt{g_s M}$, while we can estimate that the backreaction due
to the $p$ anti-branes extends over a region of order $r \simeq \sqrt{g_s p}$.
Hence the distortion of deformed conifold due to presence of the ${\overline{D3}}$
branes remains small as long as $p < \! \! < M$. We will assume that we
are in this regime.

It is now rather well understood how, in a non-trivial
flux background, $p$ D3-branes can expand to form a spherical
D5-brane (this phenomenon played an important role in the analysis of
\cite{ps}, for example). Here, due to specific form of the background
fluxes, we will need to consider the S-dual phenomenon, where the
${\overline{D3}}$ branes expand into wrapped NS 5 branes. A technical
difficulty, however, is that it is not yet known how to consistently
couple the background $B_{\it 6}$-flux (\ref{bsix}) to the matrix
Born-Infeld action of the ${\overline{D3}}$ branes. (This problem
is directly related to that of finding a Matrix theory description
of the transverse NS 5-brane.) A way around this obstacle, is to
use the S-dual description of the $\overline{D3}$ world-volume.
In \S4, we will turn things around, and view things from the
perspective of a wrapped NS 5 brane with ${\overline{D3}}$ charge
$p$ coming from a world-volume magnetic flux.

Before getting started, we need to warn the reader that
strictly speaking, since $g_s$ is assumed to be small and thus
the S-dual coupling $\tilde{g}_s = 1/g_s$ large, we are far outside
of the regime of validity of the S-dual Born-Infeld action. We
will nonetheless proceed with using it; in \S4 will find a
posteriori justification of our description when we establish a
precise match with the results obtained from the NS 5-brane
perspective.

\newsubsection{Dielectric ${\overline{D3}}$ Branes}

The worldvolume action of the $p$ anti-D3 branes,
placed at the apex of the deformed conifold and in the S-dual frame,
is given by the Born-Infeld action
\be
\label{nbi3}
S_{BI} = {\mu_3\over g_s} \int \Tr\sqrt{\det( G_\parallel + 2\pi g_s
F ) \det( Q )} - 2\pi \mu_3
\int {\rm Tr}\, {\bf i}_\Phi {\bf i}_\Phi B_{\it 6}
\ee
Here
\be
\label{qqq}
Q^i{}_j = \delta^i{}_j + {2\pi i \over g_s} [ \Phi^i, \Phi^k]
(G_{kj}+g_s C_{kj})\ .
\ee
Because we are working in the S-dual frame,
relative to the usual Born-Infeld action, we replaced $B_{\it 2}$
by $C_{\it 2}$ and $C_{\it 6}$ by $B_{\it 6}$.
The two-form $F$ here denotes the non-abelian field strength
on the ${\overline{D3}}$ brane worldvolume.

The scalar fields $\Phi$ parametrize
the transverse location $X$ of the ${\overline{D3}}$ branes, via the relation
$\Phi = 2\pi X$. By making these
matrix coordinates non-commutative, the anti D3-branes can collectively
represent a 5-dimensional brane which can be identified with the NS 5-brane.
The topology of this ``fuzzy NS 5 brane'' is ${\bf R}^4 \times S^2$, where
the two-sphere $S^2$ has an approximate radius $R$ equal to
\be
\label{rns}
R^2 \simeq  {4\pi^2 \over p} \Tr( (\Phi^i)^2).
\ee

It is instructive to look for the non-commutative solution for
$p<\!\!< M$, in which case $\Phi$ remains small relative to the
radius of curvature of the surrounding space-time and variations
in the 3-form field strengths. In this case we
may write $C_{kj} \sim {2\pi\over 3}F_{kjl}\Phi^l$,  and locally we may
approximate the  metric in the compact space by the flat metric
$G_{kj} = \delta_{kj}$.
We find that
\be
Q^{i}_{j} = \delta^{i}_{j} + {2\pi i \over g_s} [\Phi^i,\Phi_j] + i
{4\pi^2 \over 3}F_{kjl}[\Phi^i,\Phi^k] \Phi^l \ .
\ee
So we can expand
\be
\Tr\sqrt{\det( Q)} \simeq p - i
{2 \pi^2 \over 3}  F_{kjl} \Tr\Bigl( [ \Phi^k, \Phi^j] \Phi^{l} \Bigr)
-  {\pi^2\over g_s^2} \Tr\Bigl( [ \Phi^i, \Phi^j]^2\Bigr)\ .
\ee
Furthermore, we are in an imaginary self-dual flux background
where $dB_{\it 6} = -{1\over g_s} dV_4 \wedge F_{3}$.
In an imaginary $\it anti$ self-dual flux background, the cubic
terms in the full potential for the ${\overline{D3}}$ worldvolume
fields $\Phi$ would have to cancel (by a ``no-force'' condition
between anti-branes and IASD fluxes).  This logic, or alternatively,
direct calculation of the second term in (\ref{nbi3}), tells
us that the full potential coming from (\ref{nbi3}) in this
ISD flux background will be
\be
\label{mypot}
g_{s}V_{\rm ef\! f}(\Phi) \, \simeq\, \sqrt{\det(G_\parallel)}\left(\,
p - i { 4 \pi^2 \over 3} F_{kjl} \Tr\Bigl( [ \Phi^k, \Phi^j] \Phi^l \Bigr)
-  {\pi^2 \over g_s^2} \Tr\Bigl( [ \Phi^i, \Phi^j]^2 \Bigr) + \ldots\,
\right) \ .
\ee

As in \cite{Myers}, this potential has extrema away from the origin
$\Phi = 0$.  To get some intuition, let us set
$F_{kjl} = f \epsilon_{kjl}$, where we
make the approximation that the A-cycle $S^3$ is large (which
is good in the limit of large $g_s M$).
The magnitude of $f$ can be read off from the normalization of the
integrated RR flux as in (\ref{ks}), which requires
\be
f \simeq  {2 \over b_0^3 \sqrt{g^3_s M}}\ .
\ee
Demanding that ${\partial V(\Phi) \over \partial
\Phi^i} = 0$ yields the equation of motion
\be
\label{commeq}
 [[\Phi^i, \Phi^j],\Phi^j] - i g_s^2
f \epsilon_{ijk}[\Phi^j, \Phi^k] ~=~0\ .
\ee
To solve (\ref{commeq}), notice that if one takes constant
matrices $\Phi^i$
satisfying the commutation relations
\be
\label{commreln}
[\Phi^i,\Phi^j] ~=~-{i g_s^2 f }\, \epsilon_{ijk} \Phi^k ,
\ee
then (\ref{commeq}) is automatically satisfied.  But, up to rescaling
the $\Phi$, (\ref{commreln}) are just the commutation relations
which are satisfied by
a $p\! \times\! p$ dimensional matrix representation of the $SU(2)$
generators
\be
\label{sutwo}
[J^i,J^j] = 2i\epsilon_{ijk} J^k\ .
\ee
So by setting $\Phi^i = -{1\over 2} {g_s^2 f} \, J^i$, with $J^i$ the
generators of any p-dimensional $SU(2)$ representation,
we find solutions of (\ref{commreln}).

Which solution is energetically preferred?  Using the known value of the
quadratic Casimir $c_2 = {\rm Tr}((J^i)^2)$ in each $p$-dimensional
$SU(2)$ representation,
one can see that the energetically preferred solution is to take
the $p$-dimensional irreducible representation of $SU(2)$, for which one finds
\ba
\label{potev}
V_{\rm ef\! f} &\! \simeq & {\mu_3 \over g_s} \Bigl( p -
{\pi^2 \over 6}g_s^8 f^4 \,p\,(p^2 -1)\Bigr) \\[3mm]
& \simeq &  {\mu_3 p \over g_s}  \, \Bigl(\, 1 \,
- {8 \pi^2 \, (p^2-1) \over 3 \, b_0^{12} M^2}\, \Bigr)\ .
\ea
The negative term in (\ref{potev})
comes about through a competition between the
(positive) quartic term and the (negative) cubic term in
(\ref{mypot}).  The other $p$-dimensional reducible representations
occur as metastable vacua of (\ref{mypot}), where the ${\overline{D3}}$
branes have blown up to a number of ``less giant'' NS 5 branes but can
still satisfy $V_{\rm ef\! f} < p/g_s$.  These separate
5-branes all want to cluster together to form the ``most giant''
NS 5-brane, with minimal energy equal to (\ref{potev}).

It is interesting to compare the energy in the giant graviton vacuum
(\ref{potev}) to the final energy in the supersymmetric ground state,
 $V=0$. We see that $V_{\rm ef\! f} > 0$
implies that $p <\! \! < M$, which is the condition we have chosen
to impose.  In this regime, we find a self-consistent picture: the
${\overline{D3}}$ branes are driven by a perturbative instability to
expand into an NS 5 brane wrapped on an $S^2$ in the A-cycle, and must
await a non-perturbative effect to decay to the supersymmetric vacuum.
As we will discuss in \S4, the complementary analysis in
terms of the NS 5 brane worldvolume action
indicates that, for sufficiently large ${p\over M}$, the decay to a final
supersymmetric state can occur without the intermediate
metastable false vacuum. Based on the above story,
we can obtain a reasonable estimate for the onset of this classical
instability by considering the radius $R$ (given
in eqn (\ref{rns})) of the fuzzy NS 5-brane
in comparison to the radius $R_0 = b_0 \sqrt{g_s M}$ of the $S^3$
\be
\label{nsr}
R^2 
\, \simeq\, {4\pi^2 (p^2-1)\over b_0^8 M^2}\, R_0^2 
\ee
We see that there is a classical minimum only if $p/M$ is sufficiently smaller
than $b_0^4/2\pi$; otherwise the radius of the NS 5-brane will get
too close to $R_0$ and the configuration will become classically unstable.

\medskip

\sect{The NS 5-brane perspective}

In this section, we take the perspective of an
NS 5-brane moving near the tip of the conifold geometry. The
5-brane is wrapped on a two-sphere $S^2$ inside the internal $S^3$
and carries $p$ units of world-volume two-form flux which induce
 ${\overline{D3}}$ charge. As noted previously, it is a point
of some concern that the NS 5 world-volume description used
below has only limited validity for small sizes of the $S^2$.
For sufficiently large $S^2$ radius, however, the NS 5 world-volume
curvature is small compared to the string scale and one may expect that
the description as given below becomes reasonably accurate.

\subsection{Giant Graviton 5-brane}

\newcommand{\rrr}{\rho}

Consider an NS 5-brane of type IIB string theory located at an $S^2$
inside $S^3$ with radius specified by a polar angle $\psi$. The
bosonic worldvolume action reads \cite{ps}
\ba
S \is \frac{\mu_5}{g_s^2}
\int d^6\xi\, \Bigl[-\det (G_\parallel)
\det(G_\perp \! + 2\pi\alphap g_s {\cal F})\Bigr]^{1/2}
+ {\mu_5 } \int  B_{\it 6}
\ , \label{ns5act} \\[3.5mm]
& & \qquad \qquad
2\pi\alphap{\cal F}_{\it 2} = 2\pi\alphap F_{\it 2} - C_{\it
2}\ .
\ea
This action has been obtained by S-duality from that of the D5-brane.
Here $F_{\it 2} = dA$ is the two-form field strength of the world-volume gauge
field, $G_\perp$ denotes the induced metric along the internal
$S^2$, and $G_\parallel$ encodes the remaining components along the $d\psi$ and
${\bf R}^4$ directions.  Using the explicit form (\ref{apex}) of the metric,
we have
\ba
ds^2_{induced} \is  \, b_0^2 \, g_s M \Bigl[\, dx_\mu dx^\mu + 
\, d{\psi}^2 \; \;
+ \; \; \sin^2 \psi \, d\Omega_2^2\Bigr] \\[3mm]
\is
\quad \qquad \qquad
ds^2_\parallel   \qquad  \qquad +
\qquad ds^2_\perp \nonumber
\label{induced}
\ea
where (relative to eqn (\ref{apex})) we have absorbed the factor of 
${a_0/R_0}$
into $x^\mu$. (This means that from now on, all time and distance scales
in the ${\bf R}^4$ directions are measured in red-shifted string units,
or in holographic dual terminology, in terms of the dimensional transmutation
scale $\Lambda$ of the low energy gauge theory.)  We can evaluate the
following integrals over $S^2$
\ba
\label{een}
\int_{S^2} \sqrt{\det G_\perp} \is 4\pi \, b_0^2 \, g_s M \alphap \sin^2 \psi
\ , \\[2mm]
\label{twee}
\int_{S^2} C_{\it 2}(\psi) \is 4\pi \alphap
M \Bigl(\psi - {1\over 2} \sin(2\psi)\Bigr)\ ,
\\[2mm]
\label{drie}
2\pi \alphap \int_{S^2} F_{\it 2} \is \,  4 \pi^2 \alphap \, p\ .
\ea
This last equation reflects the fact that the NS 5-brane carries
${\overline{D3}}$ brane charge $p$. Combining (\ref{een})-(\ref{drie}) gives:
\be
\int_{S^2} \sqrt{\det(G_\perp \! + 2\pi\alphap g_s {\cal F})} =
4\pi^2 M g_s \alphap \, {V}_2 (\psi) 
\ee
\be
\label{vtwo}
{V}_2(\psi) =  {1\over \pi} \sqrt{b_0^4  \sin^4 \psi +
\Bigl(\pi {p\over M} \! - \! \psi\! + \! {{1\over 2}} \sin(2\psi)\Bigr)^2} .
\ee
Adding the $\mu_5 \int B_{\it 6}$ term, obtained
from eqn (\ref{bsix}), gives a total NS 5-brane action
\be
S = \int \! d^4 x\, \sqrt{-\det G_\parallel} \; {\cal L}(\psi)\ ,
\ee
\be
{\cal L}(\psi) \, = \, A_0 \Bigl( \, {V_2}(\psi)
\sqrt{ 1  - \dot{\psi}^2}
- { 1\over 2\pi}\, (2\psi - \sin 2\psi )\Bigr)\ ,
\ee
with
\be
A_0 = {4\pi^2 \mu_5 M \over g_s}  = {\mu_3 M \over g_s}.
\ee
We can use this action to introduce a canonical momentum
${P}_\psi$ conjugate to $\psi$, and write
the resulting Hamiltonian density ${\cal H}$ as
\be
{\cal H}(\psi,P_\psi) \, = \,-
{A_0\over 2\pi}\, \Bigl(2\psi - \sin 2 \psi \Bigr)
\, + \,\sqrt{ A_0^2 \Bigl({V}_2(\psi)\Bigr)^2
+ {P_\psi^2}}
\ee
which generates the time evolution of $\psi(t)$. For the moment, however,
we are just interested in finding whether there exists a static
solution corresponding to the ``giant graviton'' of \S3.

\figuur{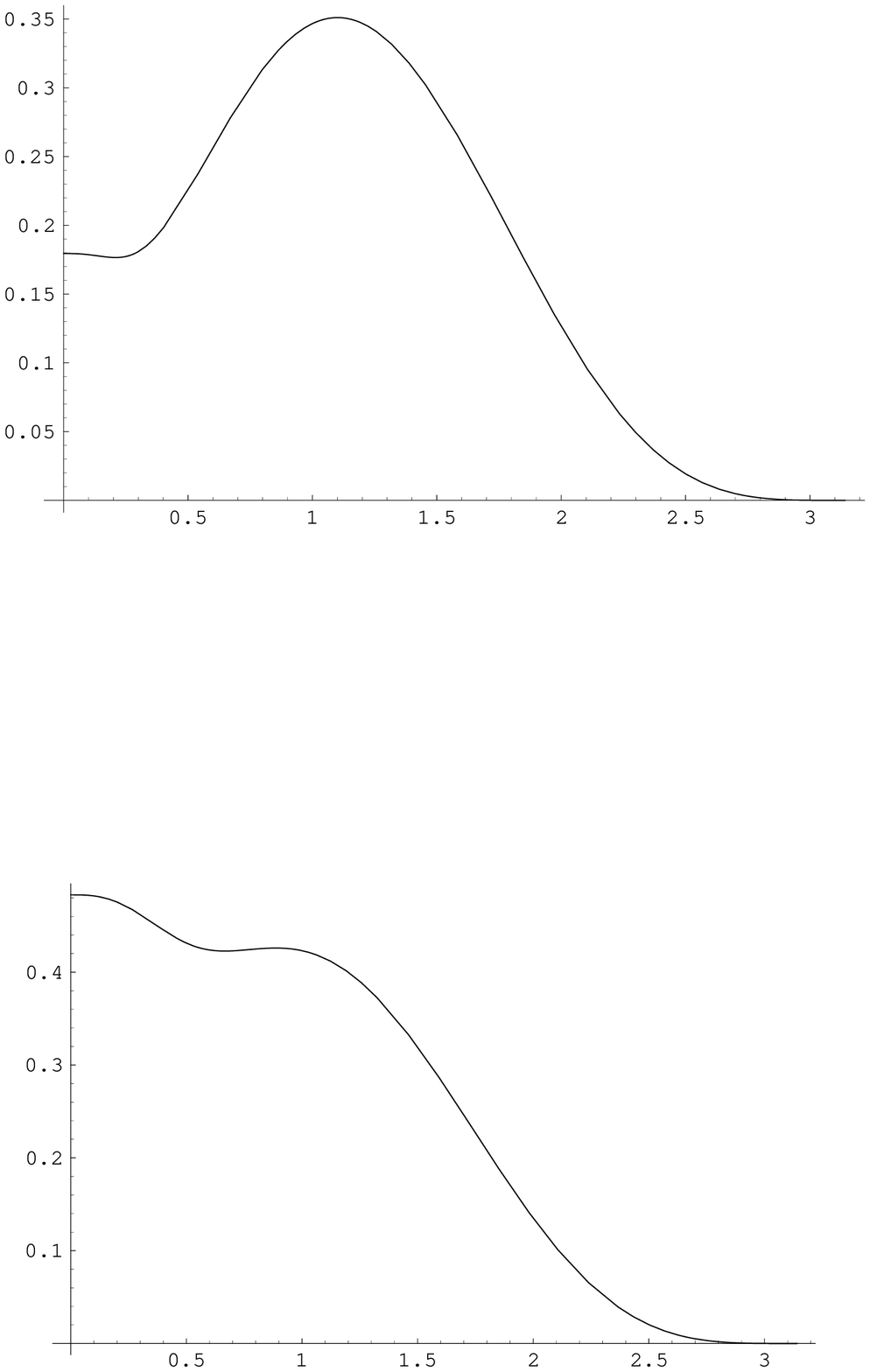}{10cm}{{\bf Fig 2:}The effective potential
$V_{\rm ef\! f}(\psi)$ for
${p\over M}\simeq {3 \%}$, showing the stable false vacuum, and
for ${p\over M}\simeq 8 \%$, with only a marginally stable minimum.}

Some useful intuition is obtained by considering the effective
potential
\ba
\label{effp1}
V_{\rm ef\!f} (\psi) & \equiv &  {\cal H}(\psi, P_\psi=0) \nonumber \\[3mm]
\is \, A_0 \, \Bigl( \, {V}_2(\psi) - {1\over 2\pi}
(2\psi - \sin 2\psi)\Bigr)
\ea
In figure 2 we have plotted this effective potential
$V_{\rm ef\! f}(\psi)$ for two different values of $p$ relative to $M$:
the top graph corresponds to ${p/ M} \simeq {3\%}$, and shows a
stable false vacuum, and the lower graph corresponds to the special
case ${p/ M}\simeq 8 \%$ at which there is only a marginally stable
intermediate minimum. For ${p/ M}> 8 \%$ the slope of
the effective potential is everywhere negative.

In both cases we can draw an interesting conclusion. In the regime
${p/ M}> 8\%$, the nonsupersymmetric configuration of
$p$ ${\overline{D3}}$ branes relaxes to the supersymmetric minimum
via a {\it classical} process: The anti-branes
cluster together to form the maximal size ``fuzzy'' NS 5-brane
which rolls down the potential until it reduces to 
$M\!-\!p$ D3-branes located at the north-pole at $\psi=\pi$.
In the regime with $p/M$ (sufficiently) smaller than $8\%$,
the branes reach a meta-stable state, the fuzzy NS-5 brane located
at the location $\psi_{min}$ for which
${\partial V_{\rm ef\!f} \over \partial \psi}(\psi_{min}) = 0$.
This configuration is classically stable, but will decay via
quantum tunneling. We will study this tunneling process in
\S 4.3. In both cases, the end result of the process is 
$M\!-\!p$ D3-branes in place of $p$ anti-D3-branes with 
the $H_{\it 3}$ flux around the B-cycle changed from $K$ to
$K\!-\!1$.

Now let us check the correspondence with the non-abelian description
of \S 3. For small values of $\psi$ we can expand
\be
\label{effapp}
V_{\rm ef\!f} (\psi) \simeq \,
A_0 \,\Bigl( {p\over M} -   {4\over 3\pi} \psi^3
+ {b_0^4 M \over 2\pi^2 p}\psi^4\Bigr)
\ee
which has a minimum at $\psi_{min} = {2\pi p\over b_0^4 M}$ equal to
\be
\label{vmin}
V_{\rm ef\!f}(\psi_{min}) \simeq  {\mu_3 p \over g_s} \, \Bigl(\, 1 \,
- {8 \pi^2 \, p^2 \over 3 \, b_0^{12} M^2}\, \Bigr),
\ee
in exact agreement with the value (\ref{potev}) found earlier. Moreover, the
size $R \simeq \psi_{min} R_0$ (with $R_0$ the $S^3$ radius) of the
NS 5 brane ``giant graviton'' exactly matches with our earlier result 
(\ref{nsr}).  

Another quantitative confirmation of the result (\ref{effp1}) for the
effective potential is that the difference in vacuum energy between the
south and north-pole is equal to twice the tension of the anti-D3 branes
\be
V_{\rm ef\! f}(0) - V_{\rm ef\! f}(\pi) = {2p\mu_3  \over g_s}.
\ee
This is the expected result. One way to understand this \cite{MaNa} is to
compare our nonsupersymmetric model with a hypothetical
situation with all $p$ anti-D3 branes replaced by $-p$ D3-branes,
i.e. branes with opposite charge and tension as $D3$-branes. This
last situation would preserve supersymmetry and would therefore
have zero vacuum energy. To change it back to our physical situation,
however, one needs to add back $p$ D3/anti-D3 pairs, with
zero charge but with total tension $2p\mu_3/g_s$. Notice, however,
that in order to obtain the true total vacuum energy, we need
to add to the result (\ref{effp1}) a term 
\be
\label{evac}
V_{\rm tot}(\psi) = V_{\rm ef\! f}(\psi) + {p\mu_3\over g_s},
\ee
so that the supersymmetric vacuum indeed has $V_{\rm tot} =0$.
This extra contribution comes from a term ${\mu_3 \over g_s}{\chi/ 24} -
\int H_3 \wedge G_3$ in the string action which, via the global
tadpole condition, adds up to $p\mu_3/g_s$.

Finally, let us return to the validity of our
description. As
mentioned, this
is a slightly problematic question, since both the S-dual D3-brane
variables and the NS 5 world-volume theory are strongly coupled.
It seems reasonable, however, that at least our main qualitative
conclusions, $(i)$ the anti-D3 branes expand to a ``fuzzy'' NS 5 brane,
$(ii)$ for small enough $p/M$, the NS 5 brane will stabilize at
some $S^2$-radius $\psi_{min}$ proportional to $p/M$, and
$(iii)$ for large enough $p/M$ the anti-D3/NS 5 configuration will
be classically unstable, will remain unchanged in a more
complete treatment. One forseeable quantitative difference,
for example, is that inclusion of the backreaction of the NS 5-brane on the
$S^3$ geometry might trigger the classical instability for smaller
values of $p/M$ than found above.

\medskip

\subsection{BPS Domain Wall}

As discussed in \S 2.3,  
one can consider supersymmetric domain walls  that interpolate
between the supersymmetric ``mesonic'' and ``baryonic'' vacuum
branches in the pure KS gauge theory with $p=0$. 
As we now show, it is possible
to write a special BPS solution to the NS 5 brane equations of
motion that describes a supersymmetric domain wall between the
two phases. Specifically, we assume that the mesonic vacuum
is such that all meson fields $(N_{ij})^{\alpha\beta}$ have the same
expectation  value. In the supergravity, this is described by the
configuration of $M$ D3-branes located at the same point on the
$S^3$, which we take to be the north-pole $\psi=\pi$.

Before we describe this domain wall solution, we note that,
for general $p$, the total ${\cal F}_{\it 2}$ flux through the
$S^2$ satisfies
\ba
2 \pi \alphap \int_{S_2} {\cal F}_{\it 2} \is 4 \pi \alphap \Bigl(\pi p
- M (\psi - {{1\over 2}} \sin(2\psi))\Bigr) \nonumber \\[3mm]
 \is -4 \pi \alphap \Bigl(\pi (M-p)
- M (\tilde{\psi} - {{1\over 2}} \sin(2\tilde{\psi}))\Bigr)
\ea
with $\tilde{\psi} = \pi - \psi$. In other words, we can think of the
${\cal F}$ background from the ``south-pole perspective'' as carrying
$p$ units of ${\overline{D3}}$ charge, or from the ``north-pole perspective''
as carrying $M\!-\! p$ units of D3-charge. Notice that this implies
that there must be $M\!-\!p$ units of $F_{\it 2}$ flux placed at the
north-pole, that is, $M-p$ D3-branes. Hence, in the special case that
$p=0$, the NS 5-brane at the north-pole represents $M$ D3 branes,
while at the south-pole it can simply shrink from view without a trace.

The domain wall solution corresponds to an NS 5-brane configuration
described by a spatial trajectory $\psi(z)$ ($z$ the coordinate
transverse to the wall) interpolating between $\psi = \pi$
and $\psi=0$. Following the same steps as above, we find that
the $z$-evolution of $\psi(z)$ is governed by the Hamilton equations
$\partial_z \psi = {\partial {\cal H}_z \over \partial P_\psi}$
and $\partial_z {P}_\psi = - {\partial {\cal H}_z \over \partial \psi}$ with
\be
{\cal H}_z(\psi,P_\psi) \, = \,-
{A_0\over 2\pi}\, \Bigl(2\psi - \sin 2 \psi \Bigr)
\, + \,\sqrt{ A_0^2 \Bigl({V}_2(\psi)\Bigr)^2
- {P_\psi^2}}
\ee
We look for a trajectory that, for large negative $z$, starts
at rest at the north-pole, i.e.,  $\psi=\pi$ and $P_\psi = 0$.
Therefore, this solution has ${\cal H}=0$. Solving for $P_\psi$ we find
\be
P_\psi = b_0^2 \sin^2 \psi.
\ee
We thus obtain the following first order equation for $\psi(z)$
\be
\partial_z \psi = -{b^2_0 \sin^2 \psi\over \psi -
{1\over 2} \sin 2\psi}.
\ee
This can be integrated to (choosing the location of
the domain wall around $z \simeq 0$)
\be
z = {\psi \cot\psi \over b^2_0 }\ .
\ee
The right-hand side covers the half-space from $z=-\infty$ to $z=1/b_0$,
where the NS 5-brane trajectory has reached the south-pole at $\psi=0$.
At this point the brane has disappeared, leaving behind the
pure flux KS solution. In the dual gauge theory, this is the baryonic
vacuum.

It is useful to think of the domain wall as an NS 5 brane wrapped
around the $S^3$ A-cycle of the conifold,
deformed near the north-pole due to the presence of the $M$ D3-branes
ending on it. From this perspective, it clearly has the
property of inducing a jump by one unit in the
$H_{\it 3}$ flux around the B-cycle. Indeed, if we consider two
such B-cycles $B(z_+)$ and $B(z_-)$ located at opposite sides of the
domain wall, their difference $B(z_+)-B(z_-)$ represents an
(otherwise contractible) 3-cycle that surrounds the NS 5-brane once.
Since the NS 5-brane acts like a magnetic source for $H_{\it 3}$ we
have
\be
\int_{B(z_+)} \! \! H_{\it 3} - \int_{B(z_-)} \! \! H_{\it 3} = 4\pi^2.
\ee

The tension of the domain wall is obtained by evaluating the classical
action of the above solution per unit time and area. We have
\be
S\; = \; \int \! dt d^2x {\sqrt{-\det G_3}} \; T_{wall}
\ee
with
\be
\label{susytension}
T_{wall} \, = \, A_0 \, b_0 \, \sqrt{g_sM} \int_0^\pi\!\! {d\psi}\; {b_0^2 \,
\sin^2 \psi \over \pi}
\, = \, {\mu_5 \, 2\pi^2 b_0^3 \, (g_s M)^{3/2}  \over g_s^2}
\ee
As expected, this result for the domain wall tension breaks up as the product
of the NS 5-brane tension $\mu_5/g_s^2$ times the volume of the $S^3$
(with radius $R_0 = b_0 \sqrt{g_sM}$) wrapped by it. From a holographic
viewpoint, this should
be compared with the formula (\ref{tension1}) obtained from the gauge theory
effective superpotential.

We should mention that the probe approximation used here is
no longer strictly 
valid near the north-pole $\psi=\pi$, since the $M$ D3-branes 
represent an appreciable stress-energy that will have a non-negligible 
effect on the background geometry. Previous experience with supersymmetric
configurations of this kind \cite{ps}, however, suggests that such
backreaction effects do not significantly alter the results for 
quantities like the domain wall tension.

\subsection{Vacuum Tunneling}

We now turn to a description of the decay of the non-supersymmetric
configuration with $p$ non-zero. This takes place via nucleation of a
bubble of supersymmetric vacuum (\ref{finalstate}) surrounded by a
spherical NS 5 domain wall which expands exponentially
as a consequence of the pressure produced by the drop in the vacuum
energy. To obtain the nucleation rate, it is standard practice to look for a
corresponding Euclidean solution. The relevant solution for us
is an NS-5 brane trajectory $\psi(R)$, where $R$ is the radial
coordinate in ${\bf R}^4$, connecting the
``giant graviton'' configuration at $\psi=\psi_{\min}$ at large $R$
to an instantonic domain wall located near some appropriate radius
$R = R_*$ at which the solution reaches the supersymmetric minimum
$\psi=\pi$.

The total action functional for such a trajectory reads
\be
S = 
B_0
\int_{R_*}^\infty \!\! dR\, R^3 \, \Bigl( \, {V_2}(\psi)
\sqrt{ 1  +  (\partial_R{\psi})^2}
+ { 1\over \pi}\, (\pi {p\over M}-\psi +{1 \over 2} \sin 2\psi )\Bigr)
\ee
with
\be
B_0 = 2\pi^2 b_0^4 \mu_3 g_s M^3.
\ee
As before, it is convenient to write the classical equations of motion
for this action in the form of Hamilton equations $\partial_R \psi =
{\partial {\cal H}_R \over \partial P_\psi}$ and $
\partial_R {P}_\psi = - {\partial {\cal H}_R \over \partial \psi}$ with
\be
{\cal H}_R(\psi,P_\psi) \, = \,-
{B_0 R^3\over 2\pi}\, \Bigl(2\psi - \sin 2 \psi \Bigr)
\, + \,\sqrt{ B_0^2\, R^6\, \Bigl({V}_2(\psi)\Bigr)^2
- {P_\psi^2}}
\ee
In figure 3 we have drawn the resulting Euclidean NS 5-brane
trajectories $\psi(R)$ for two values of $p/M$.

\figuur{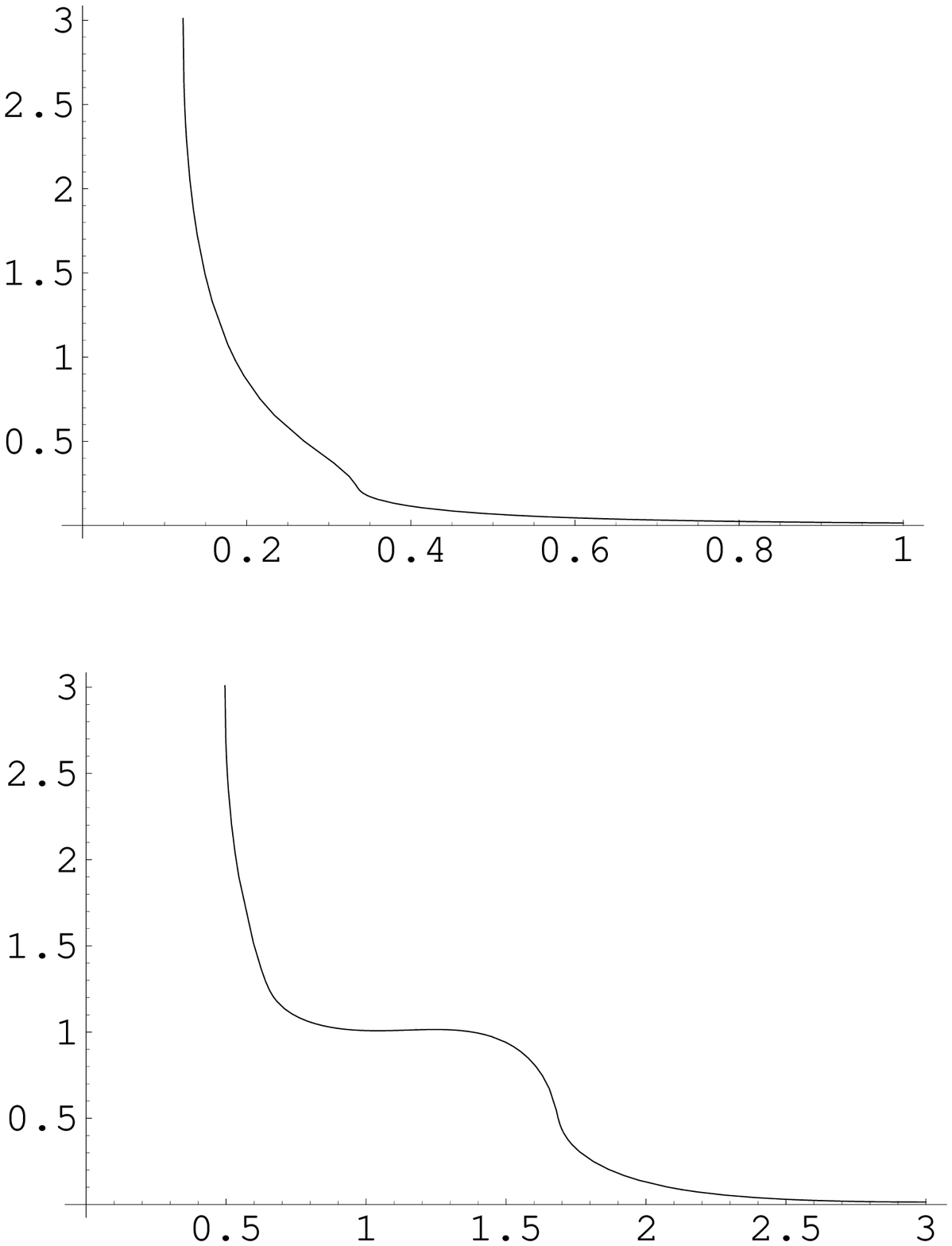}{10cm}{{\bf Fig 3:}\
The Euclidean NS 5-brane trajectory $\psi(R)$ for
two values of $p/M$: the upper trajectory corresponds to ${p \over M}
\simeq 3 \%$, and the  lower one to the near critical value
${p\over M} \simeq {8} \%$.}

In principle, we could extract the nucleation rate from the
above formulas for general $p/M$ by evaluating the total
action of the (numerically obtained) classical solutions, though
this is a laborous prospect. Instead, we can
quite easily obtain the leading order decay rate in the limit of small
$p/M$ as follows: From figure 3, the classical
trajectory for small $p$ naturally divides up into two separate
regions: it stays flat near the non-supersymmetric minimum
until coming very close to the domain wall radius $R_*$ where it
quickly moves toward the north-pole value $\psi=\pi$.  On these grounds,
we divide the total action into a contribution coming
from the non-zero vacuum energy (given in (\ref{evac}) and (\ref{vmin}))
of  the non-supersymmetric
``giant graviton'' at $R>R_*$ and a second contribution coming from the
tension of the domain wall. Using the ``thin wall approximation''
\cite{coleman} we may write
\be
S\; = \; T_{wall} \, {\rm Vol}_{{}_{\textstyle 3}}(R_*) \; - \;
V_{\rm tot}(\psi_{min})
\; {\rm Vol}_{{}_{\textstyle 4}} ( R_*)
\ee
where
\be
{\rm Vol}_{{}_{\textstyle 3}}(R_*) \, = \, 2\pi^2 
b_0^3 (g_sM)^{3/2} R_*^3 \qquad \qquad
{\rm Vol}_{{}_{\textstyle 4}}(R_*) \, = \, {1 \over 2}{\pi^2} b_0^4
(g_sM)^{2} \, R_*^4.
\ee
denote, respectively, the 3-volume of
and the 4-volume inside a 3-sphere of radius $R_*$ as measured by
the metric (\ref{induced}).

It now seems reasonable to assume that, when
$p < \! < M$, the profile of the domain wall around $R=R_*$ approaches
the supersymmetric configuration described in \S4.2.
Based on this intuition, we will set the domain wall tension $T_{wall}$
equal to its the supersymmetric value (\ref{susytension}). Taking the
leading order value $V_{\rm tot}(\psi_{min}) \simeq {2p \, \mu_3/ g_s}$
for the  energy of the false vacuum, we thus get the following
result for the classical action of the domain wall solution
with radius $R_*$
\be
S(R_*) \, = \, \pi^2 \mu_3 b_0^4 g_s M^2 \Bigl(\, M \, b_0^2 R_*^3\, - \,
p\, R_*^4\Bigr)
\ee
The two terms represent the two competing forces on the NS 5-brane domain
wall bubble: the tension pulling it inward and the outward pressure caused by
the lower energy of the supersymmetric vacuum inside the bubble. At the
critical radius
\be
\label{rcrit}
R_* = {3 M b_0^2 \over 4 p}
\ee
the two forces are balanced. Notice that, as expected, the domain wall
becomes flat in the limit that $p/M$ approaches zero.

Plugging the critical value for $R_*$ back into the action, gives the
final leading estimate of the nucleation rate
\be
\label{finalrate}
{\rm Decay~rate}~\simeq~\exp\left(
 - {27   \, b_0^{12} g_s M^6 \over 512 \pi \; p^3}\,
\right)
\ee
where we set $\mu_3=1/(2\pi)^3$. We see that the rate is very highly suppressed
in the $p<\! < M$ regime we have been considering here.

In general, the above expression gives a negligible rate in the supergravity
limit of large $M$ with $g_s M$ fixed. Note, however, that in our
derivation of (\ref{finalrate}) we have assumed that $p/M$ is far
below the critical value (about $8 \%$ in our probe approximation)
where the false vacuum becomes classically unstable. We conclude
that, by tuning $p/M$, we can make the non-supersymmetric
vacuum arbitrarily long or short-lived.

\sect{Two Dual Perspectives}

In this section we consider two dual perspectives on our model. In
particular, we propose a dual holographic description in terms
of a nonsupersymmetric, metastable vacuum in the KS field theory.
Secondly, as an additional motivation for this proposal, we summarize
how our model may be described from the type IIA point of view.

\subsection{Holographic Dual Field Theory}

For simplicity, we first consider the case that $p=1$, which
we expect to be related to the $SU(2M\! -\! 1) \times SU(M\! -\! 1)$
KS field theory. In addition to the ${\cal N}=1$ gauge multiplets, this
theory has bifundamental fields, two in the $(2M-1,{\overline{M-1}})$
representation and two in the $({\overline{2M-1}},M-1)$ representation.
So, from the perspective of the $SU(2M\! -\! 1)$,
there are $N_f\! = 2(M\! -\! 1)$ flavors, one less than the
number of colors. Hence it is no longer possible to write the color
neutral baryonic combinations ${\cal B}$ and ${\tilde{\cal B}}$,
and, as a result, the only supersymmetric vacuum of the system is the
mesonic branch (\ref{meso}). This is the holographic dual of the
stable vacuum with $M\! -\! 1$ D3-branes, the situation we expect to
land on after the decay process in the case $p=1$. The question now becomes, 
``Where do we find the other, nonsupersymmetric metastable vacuum
corresponding to a single anti-D3 brane probing the conifold?''

Though qualitatively important, for large $M$, the presence of the
single anti-D3 brane acts only as a small perturbation of the situation
with $p=0$. (For example, the effective NS 5-brane potential $V_{\rm ef\! f}$
given in (\ref{effp1})-(\ref{vtwo}) has a perfectly smooth limit
for ${p \over M} \to 0$.) It seems reasonable to assume, then, that
the nonsupersymmetric minimum for $p=1$ can be thought of as closely
related to the baryonic vacuum of the supersymmetric theory with $p=0$.
Indeed, as we have argued in \S4.2, the NS 5 domain wall that separates
it from the theory with $M-1$ D3-branes in the supersymmetric case
represents a transition between the baryonic and mesonic branch.

We will now try to use this intuition to obtain a description
of the nonsupersymmetric minimum.  We introduce, in spite of the
fact that $N_f=N_c-1$ in our case, the two ``baryonic'' superfields
\be
~{\cal B}^a~=~(A_1)^{M-1} (A_2)^{M-1},~~{\tilde{\cal B}}^a ~=~
(B_1)^{M-1} (B_2)^{M-1}\ ,
\ee
which are no longer colorless but carry a color index $a$ transforming in
the fundamental representation of $SU(2M\!-\!1)$.  The idea is
that the nonsupersymmetric theory corresponds to a false vacuum of the
$p=1$ KS gauge theory characterized by a non-zero expectation value
of these color charged ``baryon'' fields. Naturally, this will cost
energy, but it seems a reasonable assumption that (for $p/M$ very small)
this nonsupersymmetric vacuum may nonetheless be classically stable
because it is separated from the supersymmetric minimum via a
potential barrier.

To make this proposal more concrete, let us derive
the form of the superpotential of our model with ${\cal B}^a$ and
${\tilde{\cal B}}^a$ included. To this end, we start by adding to the
$SU(2M\! -\! 1) \times SU(M\! -\! 1)$ gauge theory a single pair of
scalar multiplets $A^a$ and $B^a$ (with $a$ denoting the $2M-1$ color index),
which we will then make very massive. The motivation for introducing
these extra fields is that, before we decouple them, they augment the
system to have $N_f=N_c$ so that we {\it can} introduce color neutral
baryonic fields. Define the combination $\phi_{ab} =A_aB_a$.
Now write the superpotential
\be
\label{neww}
W =  \lambda \left(N_{ij,\alpha\beta}
N_{kl}^{\alpha\beta} \right) \epsilon^{ik} \epsilon^{jl}
+ X(\det N' - \phi_{ab} {\cal B}^a\tilde{\cal B}^b -
\Lambda^{4M-2})
+ m {\rm tr} \phi
\ee
with $\det N'$ the determinant of the $(2M-1) \times (2M-1)$ meson matrix
obtained by including $A_a$ and $B_a$. Here the last term gives rise to
a mass $m$ for the extra fields $A_{a}$ and $B_a$. We can decompose
\be
det N' = \left({\rm tr} \phi + \phi_{ab} A_{i}^{\alpha, a}B_j^{\beta,b}
(N^{-1})^{ij}_{\alpha\beta} \right) \det N.
\ee
The lagrange multiplier term in (\ref{neww}) is the standard one
for an ${\cal N}=1$ gauge theory with $N_f=N_c$.
Notice that the extra field $\phi_{ab}$ does not appear in the first
term in eqn (\ref{neww}); we omit it here to avoid 
a symmetry breaking expectation value for $\phi_{ab}$. Instead, we would
like to keep $N_c$ equal to $2M\!-\!1$ after integrating
out $\phi_{ab}$. The supersymmetric $\phi_{ab}$ equations of motion now read
\be
m + X (\det N - {\rm tr} {\cal B} \tilde{\cal B}) = 0
\ee
from which we can solve for $X$, and
\be
A_{i}^{\alpha, a}B_j^{\beta,b}
(N^{-1})^{ij}_{\alpha\beta} \det N
= {\cal B}^a \tilde{\cal B}^b.
\ee
Inserting the solution for $X$ back into $W$ gives the
superpotential (with $p=1$)
\begin{equation}
\label{newsuper}
W = \lambda (N_{ij})^\alpha_\beta (N_{k\ell})_\alpha^\beta
 \eps^{ik}\eps^{j\ell} + p \left({\Lambda_{1}^{4M-p}\over
 \det_{ij,\alpha\beta} N  - \tr {\cal B} {\tilde{\cal B}}}\right)^{1\over p} \
\end{equation}
with $\Lambda_1^{4M-1} = m \Lambda^{4M-2}$. This is our proposed
superpotential of the theory with the ``baryonic'' superfields present.

In case of general $p$,
we can similarly write color charged ``baryon'' fields, which transform
in the $p$-th anti-symmetric product of the fundamental. Although
we have not done the explicit analysis in this general case, a natural
guess is that the superpotential will take the form (\ref{newsuper}).
As a primitive reasonability check, we note that the supersymmetric equation
of motion for the ``baryon'' field, $\partial_{{\cal B}} W=0$, yields
the condition that ${\cal B}^a = \tilde{\cal B}^a = 0$, so the mesonic
vacuum remains present as the only supersymmetric vacuum, according with
our expectations.

In general, without more detailed control over the dynamics, the
superpotential on its own provides at most inconclusive evidence
of the possible existence of other, nonsupersymmetric vacua of the theory.
Still, if our proposal is right, it should at least give some hint.
A general comment: Formally, the equation $\partial_{{\cal B}} W=0$
also admits one other solution, namely $\tr {\cal B} \tilde{\cal B} \to
\infty$. While it is of course dangerous to suggest that this implies the
existence of another supersymmetric vacuum, it does indicate that,
as a function of the baryon condensate, the full potential $V=|dW|^2$
of the theory will have a maximum at some intermediate scale
(which one would expect to be near $\tr {\cal B}\tilde{\cal B} \sim
\Lambda_1^{4M-4p}$). It is conceivable, therefore, that there exists another
minimum at large $\tr {\cal B} \tilde{\cal B}$.

The strongest evidence for the existence 
of the nonsupersymmetric ``baryonic'' vacuum, however, 
still comes from the supergravity analysis.
The characterization of the nonsupersymmetric model in terms of $p$
anti-D3 branes inside the conifold geometry suggests that, somewhere
in the dual field theory, there should be a hint of an (unbroken)
$SU(p)$ gauge symmetry. Indeed, our proposed dual interpretation in terms of
a phase with a non-zero condensate for $\tr{\cal B} \tilde{\cal B}$
naturally leads to a breaking of the $SU(2M\!-\!p)$ gauge symmetry
to $SU(p)$. It seems natural to identify the worldvolume theory
of the $p$ anti-D3 branes with the effective low energy description of
this $SU(p)$ sector. In the following subsection, we will find an
independent indication from the type IIA perspective that the 
nonsupersymmetric theory is described by an $SU(p)\times SU(M\!-\!p)$ 
gauge theory.

\subsection{Type IIA Brane Configurations}

The RG cascade in the KS system can also be understood via a dual
type IIA perspective \cite{KS}.  In the IIA description, one studies
a theory on D4 branes suspended between NS 5 branes.
Consider an NS 5 and an NS 5$^\prime$ brane, the first filling
out the 012345 directions, and the latter the 012389 directions
in spacetime.  Suppose them separated only along the $x^6$ direction,
compactified on a circle.   One can then stretch $N$ D4 branes
around the circle, and $M$ D4 branes on one of the two segments.
The former correspond to the $N$ D3 branes and the latter to 
fractional branes.  The resulting field theory has $SU(N+M) \times SU(N)$
gauge group and the matter stretching across the NS 5 and NS 5${^\prime}$
branes gives precisely the bi-fundamentals which arise in the
KS field theory.

In this description, the forces on the branes are not perfectly balanced
-- the NS branes bend together on the segment with the additional
fractional branes.   This corresponds to the fact that the $SU(N+M)$
gauge theory has $N_{f} = 2N$ flavors and becomes strongly coupled as
one flows to the IR.  One can move the 5$^\prime$ brane through the
NS 5 brane and around the $x^6$ circle to avoid this intersection;
this will have the effect of reducing the gauge group to $SU(N) \times
SU(N-M)$.  This is the first step in the RG cascade, and it
is
repeated until the rank of the gauge groups is low enough that
$N_{f} < N_{c}$ in one of the factors and the non-perturbative dynamics
becomes more subtle.

The difference in our setup is the addition of $p$ ${\overline{D3}}$
branes to
the conifold.  In the dual brane configuration, these should
appear as $p$ ${\overline{D4}}$ branes stretched between the NS 5 and
the NS 5$^\prime$ branes in addition to those already present
in the KS setup.  There are now (at least) two obvious options:

\noindent
$\bullet$  The $p$ anti-branes can annihilate with the D4 branes
in each segment, leaving an $SU(N\!+\!M\!-\!p) \times SU(N\!-\!p)$ 
gauge theory. This subsequently undergoes the RG cascade as described above; 
assuming $N=KM$ and $p <\! < M$, the endpoint is a supersymmetric
$SU(2M\!-\!p) \times SU(M\!-\!p)$ gauge theory.  This is, of course, the
dual gauge theory description of the final state (\ref{finalstate}).

\noindent
$\bullet$ Alternatively, one can first go through the KS RG cascade
with the D4 branes, leaving a pure $SU(M)$ gauge theory from the
D4 brane sector.  Then, including the $p$ anti-branes, one finds
a nonsupersymmetric theory with gauge group $SU(M\! -\! p) \times SU(p)$.
Special cases of this theory were discussed in \cite{Mukhi}.
This is a type IIA dual description of our nonsupersymmetric
configuration.
It would be interesting to understand the vacuum structure and
gauge symmetry breaking patterns found in
\S3\ from the IIA perspective.  The analysis in \cite{Mukhi}
finds evidence of a symmetry breaking pattern which depends
sensitively on the radius of the $x_6$ circle, but
is carried out far from the $ p <\! < M,N $
regime of interest to us here.

\newsection{Concluding Remarks}
\figuur{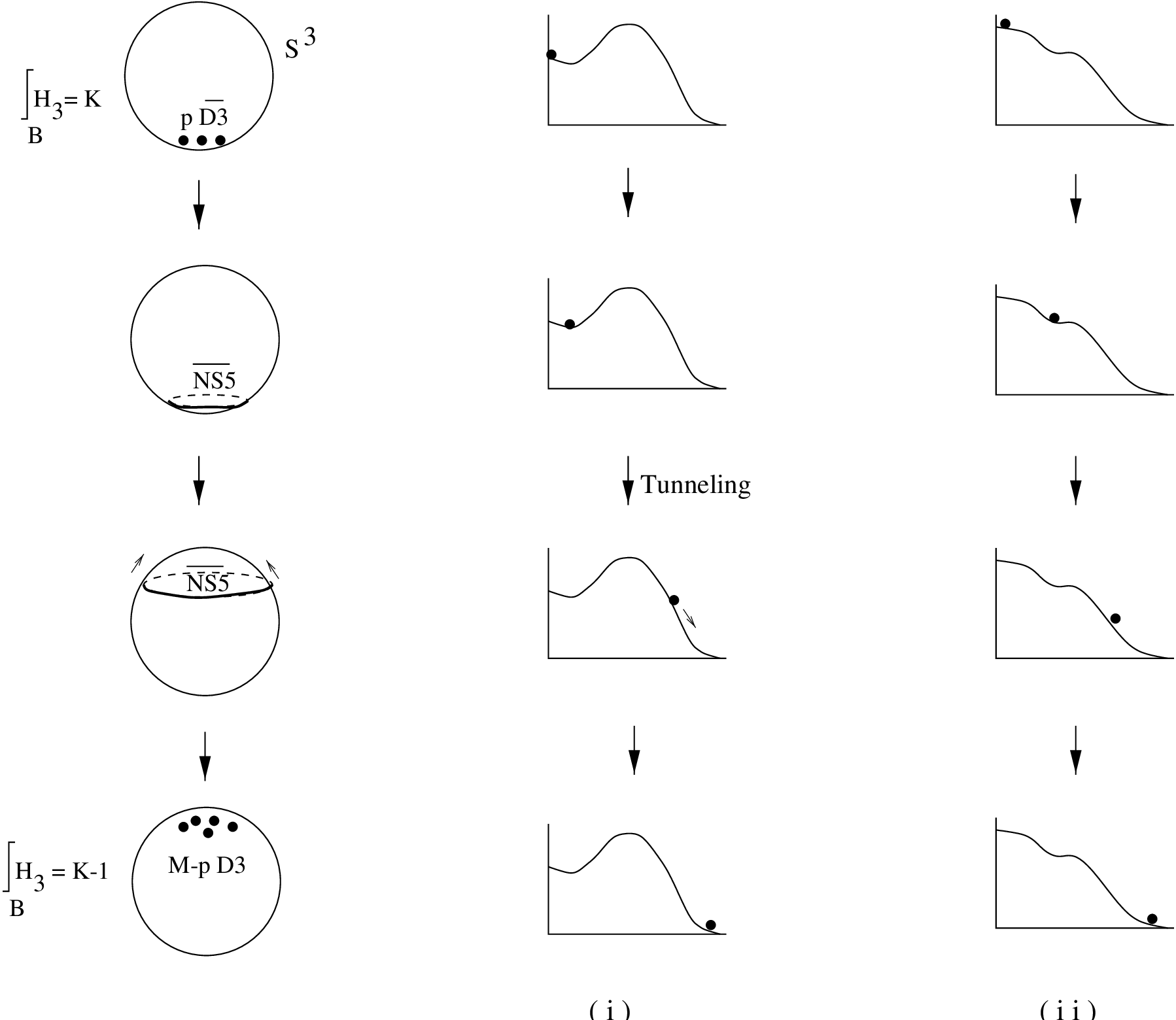}{12.5 cm}{{\bf Fig 4:}Schematic depiction of the brane/flux annihilation 
process for (i) subcritical and (ii) supercritical values for $p/M$.}
\bigskip
\bigskip

\noindent
We have found that the configuration of $p$ $\overline{D3}$ branes
probing the KS geometry
constitutes a rich system with several interesting properties.
The basic physics is apparent in Figure 4.  For
$p <\! < M$,  the system
relaxes to a metastable nonsupersymmetric vacuum but eventually
tunnels to a supersymmetric final state. This decay has a
strongly suppressed rate given in (\ref{finalrate}) and
the nonsupersymmetric vacuum can therefore be made arbitrarily long-lived.
For $p/M$ larger than some critical value (of about 8\% in our
probe approximation) the decay takes place classically.
In both cases, the decay is effected by ``brane/flux 
annihilation'' where the branes first form an NS 5-brane which
later unwinds around the $S^3$, creating $M\!-\!p$
D3 branes in the process. An important remaining problem is to 
find a supergravity solution of the nonsupersymmetric minimum
that includes the backreaction of the $p$ NS 5-branes. 

The same basic brane/flux transmutation process whereby fluxes are
traded for $D3$ branes may also provide a new perspective on 
many of the dualities currently under study as 
geometric transitions. An analogue of our microscopic description 
of this process via NS 5 brane nucleation may also play an
important role in those transitions, which encode the information 
about D-brane gauge theories in terms of dual geometries with fluxes
(see e.g. \cite{Vafa,unif}).

In most of this paper, we have restricted our attention to the infrared
physics of the model, implicitly assuming that it embedded in a
non-compact warped geometry with fixed boundary conditions in the UV.
It is an interesting question, however, to ask what happens when we embed
our model in a true string compactification with a finite volume
as constructed in \cite{fluxhier}.
In this case, the holographically dual gauge theory will be coupled
to 4-d gravity. Looking at the form of the potentials in figure 4,
it is then natural to ask what type of cosmological evolutions are
possible in this set-up.  

Besides the $\psi$ field, the string compactification will generally give rise 
to many other light moduli fields. In the basic model of 
\cite{fluxhier}, all can be made massive except for the K\"ahler 
modulus $u(x)$ controlling the overall volume of the 6d internal space $Y$.
If we ignore the backreaction due to the branes and fluxes, it is defined by
\be
\label{sfmetric}
ds^2 = g_{\mu\nu}dx^{\mu} dx^{\nu} + e^{2u} g_{i\bar j}dz^i d\bar z^{\bar j},
\ee
where $ds^2$ is the 10d string frame metric and $g_{i\bar j}$ the Ricci-flat 
metric on $Y$. Following the steps outlined in \cite{fluxhier},
one obtains for the 4d low energy effective action
\begin{equation}
\label{fourdac}
S= \frac{1}{2\kappa_{4}^2} \int\! d^{4}x\,
(-\tilde g_4)^{1/2} \Biggl( \tilde R_4 -
6 \, (\partial_\mu u\,)^2 
 -  a_0^2 \, e^{-6u} V_2(\psi) (\partial_\mu \psi\,)^2 
 + a_0^4\, e^{-12u} \, V_{\rm tot}(\psi)\, \Biggr),
\end{equation}
with $\tilde{g_4}$ the 4d Einstein frame metric, and $V_2$ and $V_{\rm tot}$ 
as in (\ref{vtwo}) and (\ref{effp1})-(\ref{evac}). 
The coefficient ${a}_0$ is the warpfactor at the location of the 
anti-D3-branes.

The dynamics of this low energy field theory is dominated by the
steep inverse-volume dependence evident in (\ref{fourdac}), which
implies that the presence of the extra energy density in the anti-branes
will quickly force the Calabi-Yau manifold to decompactify. In addition, 
it prevents the model from giving rise to any appreciable inflation.  
To stop this decompactification process, or to get an inflationary
solution, it seems that one would have to find a novel means of 
stabilizing the K\"ahler moduli (for a discussion of moduli stabilization 
in roughly this context, see e.g. \cite{fluxhier,Evafix}). 

The gravitational effect of brane nucleation processes that induce
discrete flux jumps has recently been investigated in \cite{fluxsalt}, 
as a possible dynamical mechanism for neutralizing the cosmological 
constant. In particular the set-up considered in the second reference
appears closely related to ours.

\bigskip
\bigskip

\begin{center}
{\bf Acknowledgements}
\end{center}
\medskip
We would like to thank A. Bergman, C. S. Chan, 
K. Dasgupta, S. Giddings, C. Herzog,
I. Klebanov, J. March-Russell, J. Maldacena,
S. Mukhi, P. Ouyang, J. Polchinski, M. Schulz, N. Seiberg,
M. Strassler, S. Trivedi, E. Verlinde and E. Witten for helpful discussions.
The work of J.P. and H.V. is supported in part by a National Science
Foundation grant 98-02484 and an NSF Graduate
Research Fellowship.
The work of S.K. is supported in part by a Packard Fellowship, a Sloan
Fellowship, the DOE under contract DE-AC03-76SF00515,
and National Science Foundation grant PHY-0097915.

\renewcommand{\Large}{\large}

\end{document}